\documentclass[showpacs,floats,aps,twoside,11]{revtex4}
\usepackage{amsmath,amssymb}
\usepackage{fancyhdr,lastpage}
\usepackage{graphicx}
\usepackage{epsfig}
\usepackage{setspace}
\usepackage{float}
\floatstyle{plaintop}
\restylefloat{table}
 
\begin{document}
\setcounter{page}{1}

\title{Imaginary time density functional calculation of ground states \\
for second -- row atoms using CWDVR approach}

\author{D. Naranchimeg$^1$} 
\email{Naranchimeg@must.edu.mn}
\author{L. Khenmedekh$^1$} 
\author{G. Munkhsaikhan$^1$}
\author{N. Tsogbadrakh$^2$}
\email{Tsogbadrakh@num.edu.mn}
\affiliation{$^1$Department of physics, Mongolian University of Science and Technology, Ulaanbaatar 14191, Mongolia\\
$^2$Department of physics, National University of Mongolia, Ulaanbaatar 14201, Mongolia}

\begin{abstract}
We have developed the Coulomb wave function discrete variable representation (CWDVR) method to solve the imaginary time dependent Kohn -- Sham equation on the many -- electronic second row atoms. The imaginary time dependent Kohn -- Sham equation is numerically solved using the CWDVR method. We have presented that the results of calculation for second row $Li$, $Be$, $B$, $C$, $N$, $O$ and $F$ atoms are in good agreement with other best available values using the Mathematica 7.0 programm.
\end{abstract}

\maketitle           
\thispagestyle{plain}
\setcounter{page}{1}
\goodbreak

\section{Introduction}

Numerical approach of many -- electron systems is extremely difficult computation. Density functional theory (DFT) is a computational quantum mechanical modeling method used to investigate many -- electron systems, in particular atoms, molecules, and the condensed phases \cite{ref1}. It provides a powerful alternative technique to { \it ab -- initio} wave function approach, since the electron density $\rho(\vec{r})$  possesses only three spatial dimensions no matter how large the system is. The DFT proves accurate and computationally much less expensive than usual {\it ab -- initio} wave function methods, such as a Hartree Fock method. However, the exchange -- correlation energy functional, which is a functional of the total electron density is not known exactly, and thus approximate exchange -- correlation energy functional must be used. The DFT based upon the Hohenberg -- Kohn (HK) energy functional \cite{ref2} focuses on the solution of exchange -- correlation energy and it had been used in many calculations of ground state properties an atomic system. The Kohn -- Sham equation is shown to be solved by the Coulomb wave function discrete variable representation method. Since the CWDVR method is able to treat the Coulomb singularity naturally, it is suitable for atomic systems \cite{ref3}. In our previous article, we calculated the ground state properties for noble gas atoms, such as $He$, $Ne$ and $Ar$ atoms using the Coulomb wave function discrete variable representation (CWDVR) method \cite{ref33}.   

In this paper, we present the solution of the Kohn-Sham equation on the ground state problem for the many -- electronic second row -- atoms by the CWDVR method. This paper consists of methodology and results by the CWDVR method. We show that ground state energy values calculated by the present method are in good agreement with other precise theoretical calculations.

\section{CWDVR Method} 

The DVR method has its origin in the transformation method devised by Harris et al \cite{ref4}, where it was further developed by Dickinson and Certain \cite{ref5}. Light et al. \cite{ref6} first explicitly used the DVR method as a basis representation for quantum problems, where after different types of DVR methods have found wide applications in different fields of physical and chemical problems \cite{ref7}. The DVR method gives an idea, associated basis functions are localized about discrete values of the coordinate under consideration. The DVR simplifies the evaluation of Hamiltonian matrix elements. The matrix elements of kinetic energy can also be calculated very simply and analytically in most cases \cite{ref8}. In this section, we first give a brief introduction to the DVR constructed from orthogonal polynomials and Coulomb wave functions, which will be used to solve the Kohn -- Sham equation for many -- electron atomic systems.

The DVR approach basis functions can be constructed from any complete set of orthogonal polynomials, defined in the domain with the corresponding weight function \cite{ref8}. It is known that a Gaussian quadrature can also be constructed using nonclassical polynomials. The DVR derived from the Legendre polynomials has been shown by Machtoub and Zhang \cite{ref9} to provide very precise results for the metastable states of the exotic helium atom.

An appropriate quadrature rule for the Coulomb wave function was given by Dunseath et al \cite{ref10} with explicit expressions for the weights. The time dependent single particle Kohn -- Sham equation has the form

\begin{equation}
i\frac{\partial \psi_{j} (\vec{r}, t)}{\partial t}=(\widehat{H}_0+\upsilon_{eff})\psi_{j} (\vec{r}, t), j=\overline{1, N}
\label{eqn1}
\end{equation}

Here, $\psi (\vec{r}, t)$ the single particle Kohn -- Sham orbit of $N$ electron atom, $\widehat{H}_0$ -- atomic Hamiltonian, $\upsilon_{eff}$ is the time dependent effective potential, and charge density depends on the coordinates and time and is given by
 
\begin{equation}
\rho(\vec{r}, t)=\sum^N_{j=1} |\psi_{j} (\vec{r}, t)|^2 
\label{eqn2}
\end{equation}

However, one can rewrite Eq.(\ref{eqn1}) in imaginary time $\tau$ and substitute $\tau=-i t$, $t$ being the real time, to obtain a diffusion -- type equations:
\begin{equation}
-\frac{\partial R_j (\vec{r}, t)}{\partial t}=(-\frac{1}{2} \nabla^2+\upsilon_{eff}) R_{j} (\vec{r}, t) 
\label{eqn3}
\end{equation}

The Kohn -- Sham effective local potential contains both classical and quantum potentials and can be written as:

\begin{equation}
\upsilon_{eff}[\rho; \vec{r}, t]=\frac{\delta E_{bb}}{\delta \rho}+\frac{\delta E_{ne}}{\delta \rho}+\frac{\delta E_{xc}}{\delta \rho}+\frac{\delta E_{ext}}{\delta \rho}
\label{eqn4}
\end{equation}

Here the first term is inter -- electronic Coulomb repulsion, the second is the electron -- nuclear attraction term, the third is exchange -- correlation term, and last term comes from interaction with the external field (in the present case, this interaction is zero). A simple local energy functional form has been applied for the atoms, and the exchange part can be found to be \cite{ref11},

\begin{equation}
\frac{\delta E_{x}}{\delta \rho}=\frac{\delta E^{LDA}_{x}}{\delta \rho}-\beta \bigg[\frac{\frac{4}{3}\rho^{1/3}+\frac{2}{3}\frac{r^2 \rho}{\alpha_x}}{(1+\frac{r^2 \rho^{2/3}}{\alpha_x})^2}\bigg]
\label{eqn5}
\end{equation}
\begin{equation}
\frac{\delta E^{LDA}_{x}}{\delta\rho}=-\frac{4}{3}C_x \rho^{1/3}
\label{eqn6}
\end{equation}

The simple local parameterized Wigner -- type correlation energy functional \cite{ref12} used for ground states:

\begin{equation}
E_c=-\int \frac{\rho}{a+b\cdot \rho^{-1/3}} d\vec{r}
\label{eqn7}
\end{equation}
\begin{equation}
\frac{\delta E_c}{\delta \rho}=-\frac{a+c\cdot \rho^{-1/3}}{(a+b\cdot \rho^{-1/3})^2}
\label{eqn8}
\end{equation}

where $a=9.81$, $b=21.437$, $c=28.582667$ are respectively. The solution of Eq.(\ref{eqn1}) is used split time method, for split time $\Delta t$. It can be written

\begin{equation}
R(\vec{r}, t+\Delta t)\cong e^{-\Delta t \hat{H}_0/2} e^{-\hat{V} \Delta t} e^{-\Delta t \hat{H}_0/2} R(\vec{r} ,t)
\label{eqn9}
\end{equation}

One of the main features of the DVR is that a function $R(\vec{r}, t)$ can be approximated by interpolation through the given grid points:

\begin{equation}
R(r)\cong \sum^N_{j=0} R(r_j)\cdot g_j (r)
\label{eqn10}
\end{equation}

Here: $R(r_j)$ is the interpolation function, $g_j (r)$ is the cardinal function. 

The Coulomb wave function is defined by radial grid points. Interpolation function is obtained by using the radial function that is derived from the cardinal functions. By noting that $F(r)$ is the Coulomb function, $F'(r)$ is the first derivative from F(r) at the position $r_j$, $\psi_j$ is found to be  
$\psi_j=\frac{R(r)}{F'(r)}$. The propagation in the energy space (step first in equation) can now be achieved through

\begin{equation}
e^{-\hat{H}_0 \Delta t/2} R(r)=\sum^N_{j=0} e^{-\hat{H}_0 \Delta t/2} R(r_j) g_j(r)
\label{eqn11}
\end{equation}

The cardinal functions $g_j(r)$ in (Eq.\ref{eqn10}) are given by the following expression 

\begin{equation}
g_j(r)=\frac{1}{F'(r_j)}\frac{F(r)}{r-r_j}
\label{eqn12}
\end{equation}

where the points $r_j$ ($j=1,2,...,N$) are the zeros of the Coulomb wave function $F(r)$ and $F'(r_j)$ stands for its first derivative at $r_j$ and $g_j (r)$ satisfies the cardinality condition

\begin{equation}
g_j(r_i)=\delta_{j i}.
\label{eqn13}
\end{equation}

Since the Coulomb wave functions was expressed in quadrature rule with expressions for the weight $\omega_j$, then DVR basis function $F_j(r)$ satisfies the eigenvalue for the radial Kohn -- Sham type equation: 

\begin{equation}
\hat{H} (r) \psi(r)=E \psi(r)
\label{eqn14}
\end{equation}
and 
\begin{equation}
\hat{H} (r)=-\frac{d^2}{2 d^2}+ V(r).
\label{eqn15}
\end{equation}

The DVR greatly simplifies the evaluation of Hamiltonian matrix elements. The potential matrix elements involve merely the evaluation of the interaction potential at the DVR grid points, where no integration is needed. 
The DVR basis function $f_j(r)$ is constructed from the cardinal function $g_j(r)$ as follows

\begin{equation}
f_j(r)=\frac{1}{\sqrt{\omega_j}}g_j (r),
\label{eqn16}
\end{equation}

here the weight $\omega_j$ is given in \cite{ref10}:

\begin{equation}
\omega_j \approx \frac{\pi}{a^2_j}.
\label{eqn17}
\end{equation}
\begin{equation}
a_j=F'(r_j).
\label{eqn18}
\end{equation}

The second derivative of the cardinal function $g^{''}_j (r_j)$ is given by
 
\begin{equation}
g^{''}_j (r_j)=\delta_{jk}\frac{c_{k}}{3a_{k}}-(1-\delta_{jk})\frac{a_k}{a_j}\frac{2}{(r_k-r_j)^2}, 
\label{eqn19}
\end{equation}

where $a_k$ is given by Eq.(\ref{eqn18}) and $c_k$. Here kinetic energy matrix elements $D_{ij}$ calculated using:

\begin{equation}
c_k=-a_k(2E+2Z/r), 
\label{eqn20}
\end{equation}
\begin{equation}
D_{j i}=-\delta_{ij}\frac{c_{i}}{6a_{i}}+(1-\delta_{j i})\frac{1}{(r_i-r_j)^2}, 
\label{eqn21}
\end{equation}

In the Eq.(\ref{eqn15}), to expand $R(r_j)$ in the eigenvectors of the Hamiltonian $\hat{H}_0$, we first solve the eigenvalue problem for $\hat{H}_0$ after discretization of coordinate, the differential equation for this problem can be written as:

\begin{equation}
\sum^N_{j=1}\bigg[-\frac{1}{2} D_{j i}+V(r_j) \delta_{j i}\bigg]\phi_{kj}=\varepsilon_k \phi_{kj} 
\label{eqn22} 
\end{equation}

Here $D_{j i}$ denotes the symmetrized second derivative of the cardinal function that is given as,

\begin{equation}
(D_2)_{j i}=\frac{1}{3} (E+\frac{Z}{r}),\qquad  j=i
\label{eqn23} 
\end{equation}
\begin{equation}
(D_2)_{j i}=\frac{1}{(r_i-r_j)^2},\qquad  j\neq i.
\label{eqn24} 
\end{equation}

The Eq.(\ref{eqn2}) is then numerically solved to achieve a self -- consistent set of orbitals, using the DVR method. These orbitals are used to construct various Slater determinants arising out of that particular electronic configuration and its energies computed in the usual manner. A key step in the time propagation of Eq.(\ref{eqn9}) is to construct the evolution operator $e^{-\hat{H}^0_l \Delta t/2} \cong S(l)$ through an accurate and efficient representation of $\hat{H}^0_l$. Here we extend the DVR method to achieve optimal grid discretization and an accurate solution of the eigenvalue problem of  $\hat{H}^0_l$. 

In the present work, we are particularly interested in the exploration of the improvement of the Kohn -- Sham type equation in electron structure calculation. Thus we choose the Slater wave function as our initial state at $t=0$. Note that, the differential equation for time propagation is normalized at the each time step.  Here the 152 grid points are used for the DVR discretization of the radial coordinates and $\Delta t=0.001 au.$, with 500 iteration is used in the time propagation to achieve convergence. 

\section{Calculation and Results}

In this section we present results from nonrelativistic electronic structure calculation of the ground states of $Li$, $Be$, $B$, $C$, $N$, $O$ and $F$ atoms. Here, parameters of the Coulomb wave function such as wave number and effective charges are chosen to be $k=\sqrt{2 E}=3$ and $Z1=400$ . Table \ref{tab1} summarizes the main results for mentioned atoms. The first row shows the present results. The results from the Amlan K Roy \cite{ref13} for energies for the ground states for $Li$, $Be$, $B$, $C$, $N$, $O$ and $F$ atoms are shown below the present results. The corresponding HF values from the literature are listed for comparison. For all atoms except F (mismatch 3.1\%), we found the present results of the total electronic energies are considerably match the HF values and are significantly better than the results from Amlan K Roy \cite{ref13}. 
  

\begin{table}[ht]
\begin{center}
\caption{\textbf{Calculated ground -- state properties of $Li$, $Be$, $B$, $C$, $N$, $O$ and $F$ atoms (by the unit of $au.$) along with literature data for comparison.} \label{tab1}}

\begin{tabular}{|l|l|l|l|l|l|l|l|c|}
\hline 
& & $Li$ & $Be$ & $B$ & $C$ & $N$ & $O$ & $F$ \\
\hline
$-E$ & Present work & 7.3197 & 14.582 & 24.779 & 37.9484 & 55.625 & 75.795 & 102.897\\
 & Roy [13] & 7.221 & 14.22 1& 23.964 & 36.953 & 53.407 & 73.451 & 99.734\\
 & HF [2] & 7.4332 & 14.573 & 24.529 & 37.688 & 54.400 & 74.809 & 99.400\\
$-Z/r$ & Present work & 17.054 & 33.447 & 56.728 & 88.447 & 128.915 & 179.317 & 240.433\\
 & Roy [13] & 17.115 & 34.072 & 58.143 & 88.649 & 127.326 & 176.324 & - \\
$-E_x$ & Present work & 1.752 & 2.656 & 3.732 & 5.0416 & 6.527 & 8.223 & 10.147\\
 & Roy [13] & 1.574 & 2.404 & 3.478 & 4.640 & 5.987 & 7.490 & 10.000\\
 & HF [2] & 1.781 & 2.667 & 3.744 & 5.045 & 6.596 & 8.174 & 10.020 \\
$-E_c$ & Present work & 0.0659 & 0.093 & 0.1252 & 0.1637 & 0.2058 & 0.2524 & 0.303\\
 & Roy [13] & 0.154 & 0.322 & 0.302 & 0.368 & 0.434 & 0.543 & - \\
 & HF [2]& 0.0435 & 0.094 & 0.111 & 0.1560 & 0.1890 & 0.2414 & 0324 \\
$T$ & Present work & 7.301 & 14.172 & 23.888 & 37.301 & 53.536 & 74.825 & 98.193\\
 & Roy [13] & 7.382 & 14.844 & 25.300 & 37.924 & 53.664 & 73.444 & 98.372\\
 & HF [2] & 7.433 & 14.573 & 24.529 & 37.688 & 54.401 & 74.810 & 99.410\\
\hline
\end{tabular}
\label{main}
\end{center}
\end{table}

It is satisfying that the CWDVR approach can be used to perform high precision calculation of the Kohn -- Sham type equation with the use of only a few of grid points. 

Analyses of the results for exchange and correlation energies are given in the same table separately. The results from exchange energies ($E_x$) calculations of the present calculations show a good agreement with the HF results \cite{ref2}. For the $Li$, $Be$, $B$, $C$ and $F$ atoms, the calculated exchange energy is nearly exact, while for $N$, $O$ and $F$ there is an underestimation about $1.1\%$. This indicates that the simple local exchange functional in Eq.(\ref{eqn5}) is well accurate, compare to those of Amlan J Roy \cite{ref13}. 

\begin{figure}
\includegraphics[width=14.0cm]{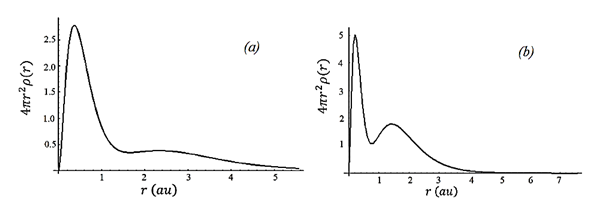}
\caption{Radial density plot of (a) $Li$ and (b) $B$ (by unit of $au.$).}
\label{fig1}
\end{figure}

\begin{figure}
\includegraphics[width=14.0cm]{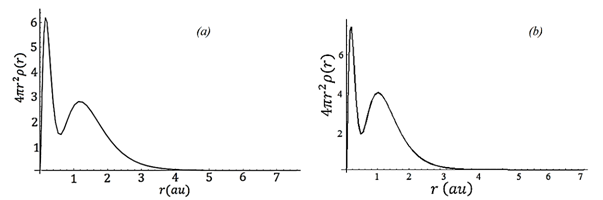}
\caption{Radial density plot of (a) $C$ and (b) $N$ (by unit of $au.$).}
\label{fig2}
\end{figure}

\begin{figure}
\includegraphics[width=12.0cm]{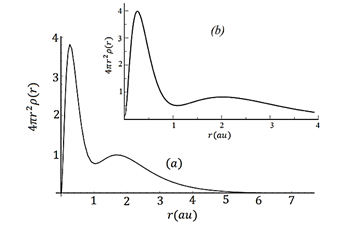}
\caption{Radial density plot of $Be$ (by unit of $au.$).}
\label{fig3}
\end{figure}

The "exact" correlation energies are considered for the$Li$, $Be$, $B$, $C$, $N$, $O$ and $F$ atoms in the Table \ref{tab1} due to the comparison with other results. The Wigner -- type correlation energy functional is likely seem to be sufficiently enough for the systems considered. For the $Be$ atom, it is nearly exact, otherwise underestimated by about $5.1-12.5\%$; the $Li$ atom is being the worst case. Compared with other generalized -- gradient approximations (GGA), Perdew's GGA \cite{ref14} correlation energy functional gives better results for $Be$, $B$ and $C$ but worse results for $Li$, $N$, $O$ and $F$. We note that the primary purpose of this work is to explore the feasibility of extending the CWDVR to the solution of the Kohn -- Sham type differential equation with imaginary time propagation. The LDA -- type $E_{xc}$ energy functionals can be easily adopted in the present CWDVR approach. Table \ref{tab1} shows that the Viral theorem is nearly satisfied for $Li$, $Be$, $B$ and $C$ atoms. The calculated kinetic energy term for the $Li$ atom is reasonably exact to HF, while for rest atoms there is an underestimation by $2.1-4.3\%$. In Figures \ref{fig1} and \ref{fig2}, the radial density plots for lithium, boroncarbon and nitrogen are presented, where HF plot is not shown. In Figure \ref{fig3}, we report the radial density plots for beryllium. The inset (a) reports the result from present calculation; the inset (b) shows the HF plot for comparison. Here, the radial density plot shape from our calculation is in good agreement with the HF plot.

\section{Conclusions}

In conclusion, we present that the nonrelativistic ground state properties of $Li$, $Be$, $B$, $C$, $N$, $O$ and $F$ atoms can be calculated by means of time -- dependent Kohn -- Sham equations and an imaginary time evolution methods. The CWDVR approach shown to be an efficient and precise solution of ground -- state energies of atoms. The calculated electronic energies are in good agreement with the HF values and are significantly better than the results in the other literatures. The approach is likely opens a road to solution of ionization and excitation states of many electron atoms.

\begin{acknowledgments}

This work was supported by the PROF2017 -- 1895 at the National University of Mongolia.
 
\end{acknowledgments}

\newpage

\end{document}